\documentclass[onecolumn,showpacs,floatfix,10pt,nofootinbib]{revtex4}
\pdfoutput=1
\usepackage{epsfig}\usepackage{float}
\RequirePackage{amssymb,amsmath}
\usepackage{makeidx}
\usepackage[latin1]{inputenc}
\usepackage{graphicx}
\usepackage{indentfirst}
\usepackage{color}
\begin{document}

\title{Target Residues Formed in the Deuteron-Induced Reaction of  Gold at Incident Energy 2.2 AGeV}
\author{A. R. Balabekyan$^a$, N. A. Demekhina$^b$, G. S. Karapetyan$^c$, D. R. Drnoyan$^d$,
V.I.Zhemenik$^d$, J.Adam$^d$, L.Zavorka$^d$, A.A.Solnyshkin$^d$, V.M.Tsoupko-Sitnikov$^d$,
J.Khushvaktov$^d$, L. Karayan$^a$, A. Deppman$^c$, V. Guimar\~aes$^c$}
\affiliation{a) Yerevan State University \\ A. Manoogian, 1, 025,
Yerevan, Armenia \\ b) Yerevan Physics Institute, Alikhanyan Brothers 2,
Yerevan 0036, Armenia\\ Joint Institute for Nuclear Research (JINR),
Flerov Laboratory of Nuclear Reactions (LNR), Joliot-Curie 6, Dubna 141980,
Moscow region Russia \\ c) Instituto de Fisica, Universidade de S\~ao Paulo \\
Rua do Matao, Travessa R 187, 05508-900 S\~ao Paulo, SP, Brazil \\ d) Joint
Institute for Nuclear Research (JINR),  Laboratory of Nuclear Problems
(LNP), Joliot-Curie 6, Dubna 141980, Moscow region Russia}

\begin{abstract}
The cross sections of 110 radioactive nuclide with
mass numbers 22 $\leq A \leq$ 198 amu from the interaction
of 2.2 GeV/nucleon  deuterons from 
the Nuclotron of the Laboratory of High Energies (LHE), Joint
Institute for Nuclear Research (JINR) at Dubna with a $^{197}$Au
target are  investigated using induced activity method. The results
including charge and mass distributions are parameterized in terms
of 3-parameter equation in order to complete the real isobaric
distribution. Using data from charge distribution total mass-yield
distribution was obtained. The analysis of the mass-yield
distribution allows to suppose existence of different channels of
the interaction such as spallation, deep spallation, fission-like
and multifragmentation processes.
\end{abstract}
\maketitle

\section*{1 Introduction}

In recent years the attention of nuclear physicists has been
directed towards understanding the mechanism of nucleus-nucleus
interactions at energies of a few GeV per nucleon. The dependence of
the formation cross section of a nuclide upon the bombarding energy
of the projectile (the excitation function) has a characteristic
shape which often can be related to the reaction mechanism. One of
the questions concerning the reaction mechanism  is  how the
projectile is interacting with target, partially or wholly.
Comparisons of cross sections of different residuals are useful in
pointing the similarities and differences between their reaction
mechanisms. Investigation of deuteron-nucleus collisions is
important since deuteron represents itself as the lightest weakly
bounded system, hence during interaction with nucleus the
characteristics of the interactions of the distinct nucleons can be derived.

In our recent works the deuteron induced reactions on separated tin
isotopes were investigated \cite{bal1}. The results obtained were
compared with proton induced reactions, which allowed to make
conclusions about the deuteron interaction mechanism. These results
represent a great interest for the improvement of theoretical
models. The experimental results can also be useful for
astrophysics over space and accelerator technology and nuclear waste
transmutation based on the accelerator-driven subcritical nuclear
power reactors.

The gold is a frequently experimentally studied target among
medium mass nuclei, which allows to track the evolution of the main
characteristics of the different reaction channels adducting to
formation of light residuals up to heavy ones. A considerable amount
of measurements was performed for proton-induced reactions on 
gold target using various experimental methods \cite{Michel,
Kaufman1, Kaufman2, Letourneau, Hashemi, Kozma1}. The main goal of
such studies was the extraction of the properties of different
channels interaction  by the analysis  of large amount of
experimental cross sections  as well as the investigation of charge
and mass distribution of the reaction products in wide range of
energy up to 300 GeV.

An extensive abundance of data now exists from the study of
intermediate and high energy heavy-ions on gold target
\cite{Gindler, Morrissey, Kozma1}. Both the counter techniques and
induced activity methods  have been successfully employed in these
studies. The calculation of the cross sections of formation of
independent products and their integration over all fragment mass
range allowed to estimate the total reaction cross section and to
compare it to the data reported in literature. This comparison allows
to specify the role of the  projectile  in the mechanism of nuclear
reaction.

Gold target is attractive because different interaction
channels are present in the experimental data and are available for
comparison. On the other hand the mono-isotope composition of the Au
element essentially facilitates the analysis of measurement data.

Nevertheless, a survey on the literature displays that there  is a
considerable lack of experimental data for the deuteron interaction
with the gold target. They are mainly devoted to the study of the
distinct reaction channels. There is a work about fission, induced
at energy 2.1 GeV, where the cross sections for binary and ternary
fission were measured using plastic detectors \cite{Rahimi}. In the
work of C. Damdinsuren $\textit{et al.}$ \cite{Kozma2}, at the
interaction of 3.65 A GeV deuterons with Au, the cross sections of
formation of spallation products were measured by direct counting of
irradiated targets on Ge(Li) spectrometers. The set of the data was
restricted by  a little amount of products, which did not allow to
make the complete analysis of the interaction. Until now, only one
study of target residues formed during the fission reactions of
deuterons with an energy of 4 GeV and gold was reported
\cite{Stoulos}. Measurements were made in the solid state nuclear
track detector technique. Here, only the total fission cross section
has been estimated and a suggestion has been made that the total
fission cross section of $^{197}$Au is the same within the accuracy
of measurements as for the fission, induced by protons with the same
total energy.

The goal of the present experiment is to provide a set of
experimental cross sections of formation of residuals in the
reactions of deuterons with Au. The experimental data obtained will
allow to estimate the contribution of different reaction channels
such as fragmentation, spallation and fission-like processes, and to
make a comparison with the earlier studies of the proton-induced
reaction.

Here we should say that in high energy nuclear reaction products can be 
produced by spallation, deep spallation,
fission, and multifragmentation processes. According to the J. Hufner 
\cite{Hufner} defined these processes in the following way:

\begin{enumerate}
\item spallation is the process in which only one heavy fragment with mass
close to the target mass $A_{T}$ is  formed (a special case of spallation is
the so-called deep spallation where $M=1$ but $A \sim \frac{2}{3}A_{T}$);
\item fission is the process in which $M=2$ and $A$ is around $A_{T}/2$;
\item multifragmentation is the process where $M>2$ and $A < 50$.
\end{enumerate}

\section*{2 Experimental Procedure}

A beam of 4.4 GeV deuteron from the Nuclotron of the VBLHEP, JINR was
used to irradiate gold target. The target was the stack of gold
foils with the size 2x2 cm$^2$. Altogether 15 foils were used with
the thickness of each target foil 39.13 mg/cm$^{2}$. The irradiation
time was 28.6 hours at ion beam  total intensity of about
$(6.43\pm0.71)\cdot 10^{12}$ deuterons. The reaction
$^{27}$Al$(d,3p2n)^{24}$Na with cross section of $15.25\pm1.5$ mb
\cite{Banaigs} for beam monitoring was used. 
The $\gamma$-rays from the decay of of residual nuclei formed in the target were measured, in an off-line analysis, with High purity Germanium (HpGe) detector with 28\% relative efficiency and an energy resolution of 2 keV ($^{60}$Co at 1332 keV). The energy-dependent efficiency of the HpGe detectors was measured with
standard calibration sources of  $^{54}$Mn, $^{57;60}$Co,
$^{137}$Cs, $^{154}$Eu, $^{152}$Eu, and $^{133}$Ba. The $\gamma$
spectra were evaluated with the code package DEIMOS32 \cite{frana}.
The residual radioactive nuclei were identified by the energy and
intensity of characteristic $\gamma$-lines and by the respective
half-lives of nucleus. Nuclear properties, used for identification
of observed isotopes, were taken from literature \cite{Firestone}.
The half-lives of identified isotopes were within the range of 15
min and 1 yr. The error in determining cross sections depended on
the following factors: the statistical significance of experimental
results ($\leq$ 2-3\%), the accuracy in measuring the target
thickness and the accuracy of tabular data on nuclear constants
($\leq$ 3\%), and the errors in determining the detector efficiency
with allowance for the accuracy in calculating its energy dependence
($\leq$ 10\%).

The fragment production cross sections are usually considered as an
independent yield (I) in the absence of a parent isotope (which may
give a contribution in measured cross section via $\beta^{\pm}$-
decays) and are determined by using the following equation:

\begin{eqnarray}
\hspace{-0.2cm}\sigma=\frac{\Delta{N}\;\lambda}{N_{d}\,N_{n}\,k\,\epsilon\,\eta\,(1-\exp{(-\lambda
t_{1})})\exp{(-\lambda t_{2})}(1-\exp{(-\lambda t_{3})})}\label{g1}
\end{eqnarray}
\noindent where $\sigma$ is the cross section of the reaction
fragment production (mb); $\Delta{N}$ is the area under the
photopeak; $N_{d}$ is the deuteron beam intensity (min$^{-1}$);
$N_{n}$ is the number of target nuclei (in 1/cm$^{2}$ units);
$t_{1}$ is the irradiation time; $t_{2}$ is the time of exposure
between the end of the irradiation and the beginning of the
measurement; $t_{3}$ is the measurement time; $\lambda$ is the decay
constant (min$^{-1}$); $\eta$ is  the intensity of
$\gamma$-transitions; $k$ is the total coefficient of $\gamma$-ray
absorption in target and detector materials, and $\epsilon$ is the
$\gamma$-ray detection efficiency.

In the case where the cross section of a given isotope includes a
contribution from the $\beta^{\pm}$-decay of neighboring unstable
isobars, the cross section calculation becomes more complicated
\cite{Baba}. If the formation cross section of the parent isotope is
known from experimental data, or if it can be estimated on the basis
of other sources, the independent cross sections of daughter nuclei
can be calculated by the relation:

\begin{widetext}

\begin{eqnarray}
\sigma_{B}=&&\frac{\lambda_{B}}{(1-\exp{(-\lambda_{B}t_{1})})\,\exp{(-\lambda_{B}t_{2})}
(\,1-\exp{(-\lambda_{B} t_{3})})}\times\nonumber\\
&&\hspace*{-1.5cm}\left.\Biggl[\frac{(\Delta{N})_{AB}}{N_{d}\,N_{n}\,k\,\epsilon\,\eta}-\sigma_{A}\,f_{AB}\,
\frac{\lambda_{A}\,\lambda_{B}}{\lambda_{B}-\lambda_{A}}
\Biggl(\frac{(1-\exp{(-\lambda_{A} t_{1})})\,\exp{(-\lambda_{A} t_{2})}\,(1-\exp{(-\lambda_{A} t_{3})})}
{\lambda^{2}_{A}}\right.\nonumber\\&&\left.\qquad\qquad\qquad\qquad\qquad\qquad\quad\qquad\qquad\qquad-\frac{(1-\exp{(-\lambda_{B} t_{1})})\,\exp{(-\lambda_{B}
t_{2})}\,(1-\exp{(-\lambda_{B} t_{3})})}{\lambda^{2}_{B}}\Biggr)\right.\Biggr],
\end{eqnarray}
\end{widetext}

\noindent where the subscripts $A$ and $B$ label variables referring
to, respectively, the parent and the daughter nucleus; the
coefficient $f_{AB}$ specifies the fraction of $A$ nuclei decaying
to a $B$ nucleus ($f_{AB}=1$, when the contribution from the
$\beta$-decay corresponds 100\%); and $(\Delta{N})_{AB}$ is the
total photopeak area associated with the decays of the daughter and
parent isotopes. The effect of the precursor can be negligible in
some limiting cases: where the half-life of the parent nucleus is
very long, or in the case where its contribution is very small. In
the case when parent and daughter isotopes can not be separated
experimentally, the calculated cross sections are classified as
cumulative ones (C). It should be mentioned that the use of
induced-activity method imposes several restrictions on the
registration of the reaction products. For example, it is impossible
to measure a stable and very short-lived, very long-lived isotopes.

\section*{3 Results and Discussion}

The experimental cross sections of reaction fragment production  in
the mass range of 22 $\leq A \leq$ 198 amu are presented in Table I.
Here, the types of cross sections (I-independent and C-cumulative)
and the type of decay ($\beta^-$and EC) are also shown. As we can
see from Table I most of residuals are the nuclei with $\beta^+$
decay. It may be due to the fact that the average multiplicity of
neutrons emitted from nucleus is a few times higher than the average
multiplicity of protons in nuclear reactions \cite{Cugnon}. 

In order to obtain a complete picture of the charge and mass
distributions of the reaction products, it is necessary to estimate
the cross sections of isotopes unmeasurable by the induced-activity
method. For this purpose the charge distribution curve (i. e., the
variation of cross section with $Z$ value at constant $A$)
constructed on the base of independent cross section of the reaction
products can be used. In the present work the analysis of the charge
distributions was obtained using the expression from
\cite{Kaufman1}:

\begin{eqnarray}
\sigma(Z, A)=\sigma(A)\exp\left(-R\left|Z-SA+TA^{2}\right|^{3/2}\right),
\end{eqnarray}
where $\sigma(Z, A)$ is the independent cross section for a given
nuclide  with atomic charge $Z$ and a mass number $A$;
$\sigma(A)$ is the total isobaric cross section of the mass
$A$, and the parameters $R$, $S$ and $T$ were fitted to the data
from the spallation mass region to $A=40$ amu. Parameter $R$ defines
the width of the charge distribution and parameters $S$
and $T$ define the most probable charge ($Z_{p}$) for a given isobar
chain $A$.

In order to uniquely specify the variables $R$, $S$ and $T$, it
will be necessary to measure more than four independent yield cross
sections for each isobar. In case of shortage of the experimental
data on the independent cross section another assumption was used
for the construction of the charge distributions. This assumption is
based on the similarity of the charge distribution curves for neighboring 
isobaric chains observed previously in different works \cite{Kudo,
Branquihno}. Thus cross sections of radionuclide from a limited
mass range can be used to determine a single charge distribution
curve. Taking into account the parent isotope cross section from the
calculated isobaric-yield distribution using Eq. 2 the measured cross
sections were adjusted on the precursor feeding where necessary.

During fitting procedure it was found that the values of $R$ and $S$
were unchanged for all mass range of the reaction products, but the
parameter $T$ was larger for the spallation fragments. This means
that the width of the charge distribution at a given mass number is
the same for all range of product mass number, but the most probable
charge is smaller for lighter mass chains (i. e., ($Z-Z_{p}$)
difference is more neutron excessive). The values of these
parameters for the 4.4-GeV deuterons are

\begin{eqnarray}
R=30A^{-0.79},   \qquad S=0.47, \qquad  T=2.3\times10^{-4}
\end{eqnarray}

Using these values and the expression (3) the total isobaric cross
section ($\sigma_{A}$) was calculated for every mass number in the
mass range $40 \leq A \leq 130$ amu. The same values of $R$ and $S$
were used for the spallation mass range ($ A > 130$) but with
parameter $T=3.2\times10^{-4}$, which reflects the shift in the most
probable charge to a smaller value than others mass ranges. The
summing of all isobaric yields gives the total mass yield for the
interaction channels in the given mass range of the residual nuclei.
It should be mentioned that in frames of intranuclear cascade model
the intermediate nuclear state with higher excitation energies
results in a large number of the evaporated nucleons. Hence, a set
of neutron-deficient nuclei are formed as a result of spallation,
deep spallation and fission-like processes, which we have
observed \cite{Maslov, Duijvestijn}. The large number of the neutron
evaporation is connected to the high excitation energy of after
cascaded nucleus. Displacement of the charge distribution curves for
residuals with $A < 130$ to the neutron deficit side can be a result
of contribution of the higher excited after cascaded nuclear states.

Fig. 1(a, b, c) shows the calculated partial cross sections (the
ratio of cross section of fragment production to the total cross
section with given mass number), as well as their Gaussian charge
distribution, for different isobaric chains as a function of the
difference ($Z_{p} - Z$). An interesting feature is that the width
of charge distribution of the present work is the same as for the
system $^{197}$Au+p at the proton energy 1.0-3.0 GeV, but in this
work the center of charge distribution for the mass $A \leq 130$ amu
shifted towards the neutron-deficient side of the valley of
beta-stability. There is also a good agreement with the parameters
$S=0.477$ and $T=2.4\times10^{-4}$ for the system $^{2}$H+$^{181}$Ta
at the energy of deuterons 7.3 GeV \cite{Kozma3}. It can been
reposed that the charge distributions of the fragments are mainly
determined by the properties of residual nuclei and don't depend on
the way of their formation.

The isobaric yields obtained in this manner are shown as the filled
squares in Fig. 2. The smooth curve on the Fig. 2 indicates the best
fit for the mass number $24 \leq A \leq 198$ amu based on the
experimental data with the inclusion of contribution from different
processes such as spallation, deep spallation or fission-like binary
decay, and fragmentation or multifragmentation. The analysis of the
cross sections distribution, depending on the product mass number or
a number of the emitted nucleons, allows to make a definite
assumption about mechanism of their formation. It can distinguish
the regions of the specific channels of the fragments formation.

The residuals like $^{196}$Au and $^{198}$Au can be formed in the
processes stripping or transition of one neutron. At the energies
investigated in present work (2.2.GeV/nucleon) the nuclear
interaction proceeds via surface collision with small energy
transition. The experimentally determined cross section for deuterons with nuclei can be
about ~1.7 times more with respect to the proton-induced reaction. It can be seen from Table I that the
probability of the neutron evaporation several times more than
neutron transfer or other processes bring to the increasing mass
number of the target nucleus. The cross section of neutron
production in deuteron-nucleus reaction in the energy range of
about GeV is larger than in proton-nucleus reaction because of
stripping reaction \cite{Cugnon}. The inelastic cross section of
neutron-nuclear reactions in energy range several GeV is about $1-2$
b \cite{Barashenkov}. Therefore the reaction $(n,2n)$ from the
stripping neutrons can give the contribution in
$^{197}Au(d,X)^{196}Au$ reaction and can increase the cross section
of our experimental data.

The region $A > 131$ amu is connected to the spallation process
with a characteristic decrease of the cross sections of the product
formation on the number of the emitted nucleons. 

The mass range 60 $\leq A \leq$ 120 amu includes different
channels of interaction indistinguishable in activation
measurements. In our previous work it was  marked, that in this
region the approximately constant value of the  excitation energy is
preserved. It can be proposed that all the above mentioned
processes (deep spallation, fission, fragmentation) contribute to
the formation of residuals. In this range the energy transfer to the
after cascaded remnants promotes the new reaction channel opening.

The intermediate mass fragments (IMFs) ($40 < A < 60$) represent the
group of fragments for which one can not say unequivocally that they
are formed during the fragmentation mechanism. One of the possible
ways of the origin of IMFs is that they represent the products in
pair with $A \sim 120-130$ amu. Or, for instance, deep spallation
may be accompanied not only by emission of nucleons and light charge
particles (with $Z \leq 2$) but emission of IMFs is not excluded
either. An alternative explanation of the origin of the products
with $A = 40-60$ amu was suggested by the intranuclear cascade model
\cite{Yariv}, according to which these fragments are the result of
the disintegration of highly excited residual nucleus with $A \sim
185$.

In region of masses less than 40 amu can be a factor of phase
transition at interaction that is connected with  increasing
excitation and transition to other kind processes as
multifragmentation \cite{Wolfgang}.

Integration of the mass-yield curves over mass number gives the
cross section for the production of target residues 2.20$\pm$0.44 b.
We have chosen a lower limit of 40 mass units because the
multiplicity of fragments with masses smaller than $\sim$ 40 is
unknown. These products may arise from interaction where another
heavy fragment also survives and had thus already been counted.

\newpage
\begin{center}
Table I. The cross sections of nuclides formed by the
reaction of 4.4 GeV deuterons with ${}^{197}$Au. Independent cross sections
are indicated by (I); others are cumulative (C).
\end{center}
\begin{center}
\begin{tabular}{||c||c||c||c||c||c||} \hline\hline
Element &Reac.(decay)&$\sigma, mb$ & Element &Reac.(decay)&$\sigma,
mb$\\ \hline $^{24}$Na& C ($\beta^-$) &8.02$\pm$0.80&$^{117m}$Sn& C
($IT$)&0.23$\pm$0.02\\ \hline $^{28}$Mg& C ($\beta^-$)
&2.30$\pm$0.66&$^{124}$Sb& I ($\beta^-$)&2.80$\pm$0.60\\ \hline
$^{42}$K& C ($\beta^-$) &3.34$\pm$0.13&$^{126}$Sb& C
($\beta^-$)&2.40$\pm$0.46\\
\hline $^{43}$K& C ($\beta^-$) &4.98$\pm$1.70&$^{119}$Te& C
($EC$)&0.66$\pm$0.09\\ \hline $^{47}$Ca& C ($\beta^-$)
&0.70$\pm$0.16&$^{132}$Te& C
($\beta^-$)&2.93$\pm$0.88\\
\hline $^{44}$Sc& I ($EC$) &1.76$\pm$0.70&$^{124}$I& I
($EC$)&0.46$\pm$0.03\\ \hline $^{44m}$Sc& I ($EC$)
&0.45$\pm$0.16&$^{126}$I& I ($EC, \beta^-$)&2.02$\pm$0.34\\ \hline
$^{46}$Sc& I ($\beta^-$) &4.05$\pm$0.46&$^{133}$I& C
($\beta^-$)&1.22$\pm$0.16\\ \hline $^{48}$Sc& I ($\beta^-$)
&1.71$\pm$0.71&$^{127}$Xe& C ($EC$)&10.92$\pm$1.20\\ \hline
$^{48}$V& C ($EC$) &1.14$\pm$0.09&$^{131}$Ba& C
($EC$)&14.94$\pm$1.23\\
\hline $^{52}$Mn& C ($EC$) &0.66$\pm$0.09&$^{140}$Ba& C
($\beta^-$)&0.79$\pm$0.08\\ \hline $^{54}$Mn& I ($EC$)
&3.59$\pm$0.69&$^{139}$Ce& C ($EC$)&11.71$\pm$0.95\\ \hline
$^{59}$Fe& C ($\beta^-$) &1.81$\pm$0.36&$^{143}$Ce& C
($\beta^-$)&7.37$\pm$1.60\\ \hline $^{55}$Co& C ($EC$)
&2.26$\pm$0.72&$^{143}$Pm& C ($\beta^-$)&12.94$\pm$0.96\\
\hline $^{56}$Co& C ($EC$) &0.34$\pm$0.10&$^{144}$Pm& C
($EC$)&1.54$\pm$0.14\\ \hline $^{58}$Co& I ($EC$)
&3.77$\pm$0.15&$^{148m}$Pm& I ($\beta^-$)&1.10$\pm$0.29\\ \hline
$^{60}$Co& C
($\beta^-$)&3.22$\pm$0.47&$^{145}$Eu& C ($EC$)&14.36$\pm$2.00\\
\hline
$^{65}$Zn& C ($EC$)&4.29$\pm$0.33&$^{146}$Eu& I ($EC$)&18.12$\pm$1.80\\
\hline $^{72}$Zn& C ($\beta^-$)&0.16$\pm$0.09&$^{147}$Eu& I
($EC$)&34.00$\pm$3.00\\ \hline $^{71}$As& C
($EC$)&0.30$\pm$0.04&$^{148}$Eu& I
($EC$)&1.09$\pm$0.17\\
\hline $^{72}$As& C ($EC$)&3.40$\pm$0.22&$^{149}$Eu& C
($EC$)&19.16$\pm$1.42\\ \hline $^{74}$As& I
($EC$)&3.02$\pm$0.24&$^{146}$Gd& C ($EC$)&13.24$\pm$2.08\\ \hline
$^{75}$Se& C ($EC$)&4.82$\pm$0.12&$^{147}$Gd& I
($EC$)&6.23$\pm$1.20\\ \hline $^{77}$Br& C
($EC$)&2.18$\pm$0.42&$^{149}$Gd& I ($EC$)&17.86$\pm$0.58\\ \hline
$^{82}$Br& I
($\beta^-$)&1.68$\pm$0.05&$^{153}$Gd& I ($EC$)&9.70$\pm$0.90\\
\hline $^{81}$Rb& C ($EC$)&5.92$\pm$0.66&$^{147}$Tb& C
($EC$)&2.99$\pm$0.50\\
\hline $^{83}$Rb& I ($EC$)&7.17$\pm$0.90&$^{149}$Tb& C
($EC$)&5.39$\pm$0.97\\
\hline $^{84}$Rb& I ($EC$)&3.20$\pm$0.30&$^{153}$Tb& C
($EC$)&4.01$\pm$0.97\\ \hline $^{83}$Sr& C
($EC$)&4.04$\pm$0.07&$^{160}$Tb& I ($\beta^-$)&8.94$\pm$0.80\\
\hline $^{85}$Sr& C ($EC$)&10.30$\pm$2.10&$^{167}$Tm& C
($EC$)&21.80$\pm$3.20\\ \hline $^{86}$Y& C
($EC$)&6.21$\pm$0.88&$^{168}$Tm& I ($EC, \beta^-$)&2.62$\pm$0.18\\
\hline $^{87}$Y& C ($EC$)&8.93$\pm$1.02&$^{171}$Lu& C ($EC$)&
26.75$\pm$3.00\\ \hline $^{88}$Y& C ($EC$)&6.01$\pm$0.80&$^{172}$Lu&
C ($EC$)&
4.16$\pm$0.42\\
\hline $^{88}$Zr& C ($EC$)&9.44$\pm$0.50&$^{173}$Lu& C ($EC$)&
25.16$\pm$1.80\\ \hline $^{89}$Zr& C
($EC$)&7.26$\pm$0.90&$^{177m}$Lu& C ($\beta^-$)& 1.36$\pm$0.30\\
\hline $^{95}$Zr& C
($\beta^-$)&0.7$\pm$0.38&$^{175}$Hf& C ($EC$)& 27.14$\pm$3.00\\
\hline $^{90}$Nb& C ($EC$)&4.80$\pm$0.94&$^{181}$Hf& C ($\beta^-$)&
2.41$\pm$0.24\\ \hline $^{92m}$Nb& I ($EC,
\beta^-$)&0.38$\pm$0.04&$^{182}$Ta& C ($\beta^-$)& 2.75$\pm$0.75\\
\hline $^{95}$Nb& I ($\beta^-$)&1.09$\pm$0.30&$^{182}$Re& C ($EC$)&
48.97$\pm$9.40\\ \hline $^{95m}$Nb& C
($\beta^-$)&0.35$\pm$0.03&$^{184g}$Re& I ($EC$)&
0.92$\pm$0.18\\
\hline $^{93}$Tc& C ($EC$)&3.46$\pm$0.42&$^{184m}$Re& I ($EC$)&
4.09$\pm$0.25\\ \hline
$^{94}$Tc& C ($EC$)&2.67$\pm$0.34&$^{185}$Os&I ($EC$)& 28.95$\pm$5.40\\
\hline $^{96}$Tc& I ($EC$)&2.25$\pm$0.30&$^{185}$Ir& C ($EC$)&
26.88$\pm$2.18\\ \hline \hline
\end{tabular}
\end{center}
\vspace{2cm}

\begin{center}
Continue of the Table 1.
\end{center}
\begin{center}
\begin{tabular}{||c||c||c||c||c||c||} \hline\hline
$^{103}$Ru& C ($\beta^-$)&1.32$\pm$0.13&$^{190}$Ir& I ($EC$)&
4.09$\pm$0.80\\ \hline $^{99}$Rh& C ($EC$)
&4.92$\pm$0.07&$^{192}$Ir& I ($EC, \beta^-$)& 3.12$\pm$0.70\\ \hline
$^{100}$Rh& I
($EC$)&2.13$\pm$0.12&$^{194m}$Ir& C($\beta^-$) & 1.13$\pm$0.10\\
\hline $^{102}$Rh& I ($EC,
\beta^-$)&3.85$\pm$0.60&$^{188}$Pt& C ($EC$)& 27.35$\pm$8.00\\
\hline $^{102m}$Rh& I ($EC, \beta^-$)&13.15$\pm$1.50&$^{191}$Pt& C
($EC$)& 14.95$\pm$3.50\\ \hline $^{100}$Pd& C
($EC$)&1.31$\pm$0.36&$^{194}$Au& C ($EC$)& 26.61$\pm$2.00\\ \hline
$^{105}$Ag& C ($EC$)&8.06$\pm$1.10&$^{195}$Au& C ($EC$)&
15.41$\pm$2.60\\ \hline $^{106m}$Ag& I ($EC,
\beta^-$)&2.31$\pm$0.40&$^{196}$Au& I ($EC,
\beta^-$)&140.82$\pm$12.82\\ \hline $^{110m}$Ag& I ($EC,
\beta^-$)&0.32$\pm$0.02&$^{198g}$Au& I ($\beta^-$)& 1.56$\pm$0.25\\
\hline $^{111}$In& C ($EC$)&8.03$\pm$1.10&$^{198m}$Au& I
($\beta^-$)& 7.41$\pm$0.50\\ \hline $^{113}$Sn& C
$EC$)&17.43$\pm$1.98& & &\\ \hline \hline
\end{tabular}
\end{center}
\vspace{2cm}

For comparison we have calculated the total reaction cross section
using the hard sphere model \cite{Bradt} for nucleus-nuclear
interactions. This model refers to the overlap of the two sharp
spheres forms of interacting nuclei and the total reaction cross
section is presented in terms of a two-parameters expression:

\begin{eqnarray}
\sigma_{R}=\pi r^{2}_{0}(A^{1/3}_{T}+A^{1/3}_{p}-b_{Tp})^{2}    fm^{2},
\end{eqnarray}
where $A_{T}$ and $A_{p}$ are the mass numbers of the target and
projectile nuclei, respectively; $r_{0}$, is the constant of
proportionality in the expression of geometrical nuclear radius
$r_{i}=r_{0}A^{1/3}_{i}$ and $b_{Tp}$ is the overlap parameter. The
quantity $b_{Tp}$ is equal to $\Delta r/r_{0}$ where $\Delta r$ is
the geometrical overlap between the colliding nuclei.

Because the total reaction cross section is known from our
experimental data, the impact parameter $b$ can be estimated using
the above mentioned expression. From estimated overlap parameter
$b_{Tp}=0.97$ fm in the present experiment the mean value of the
impact parameter $b$ equals to 8.37 fm was obtained.

\newpage
\begin{figure*}[h!]
\includegraphics[width=16cm]{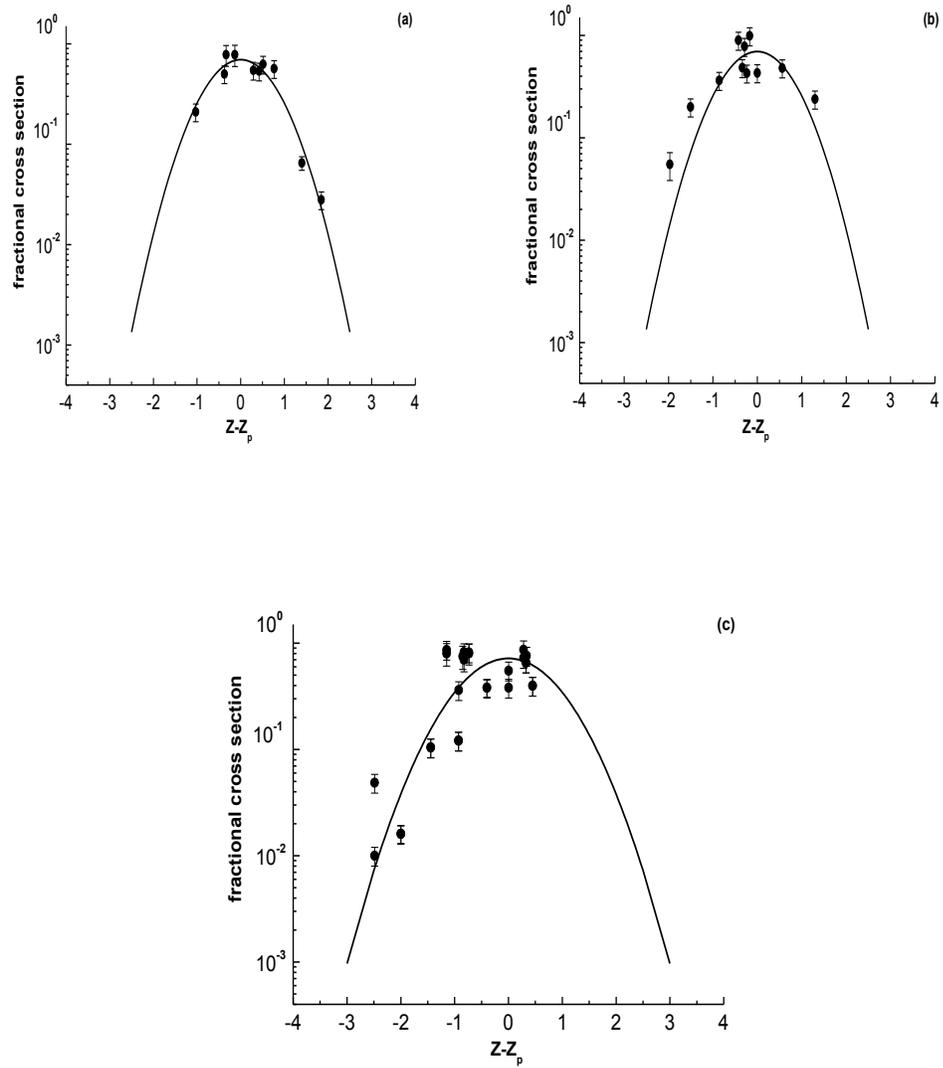}
\caption{\small The charge distributions of the sobaric chains
in the mass regions: (a)=42-65; (b)=83-127; (c)=133-198 for the
4.4 GeV deuteron-induced reaction on $^{197}$Au;
$\bullet$-experimental data; solid line is fit by expression (3) .}
\end{figure*}

\newpage
\begin{figure*}[h!]
\includegraphics[width=16cm]{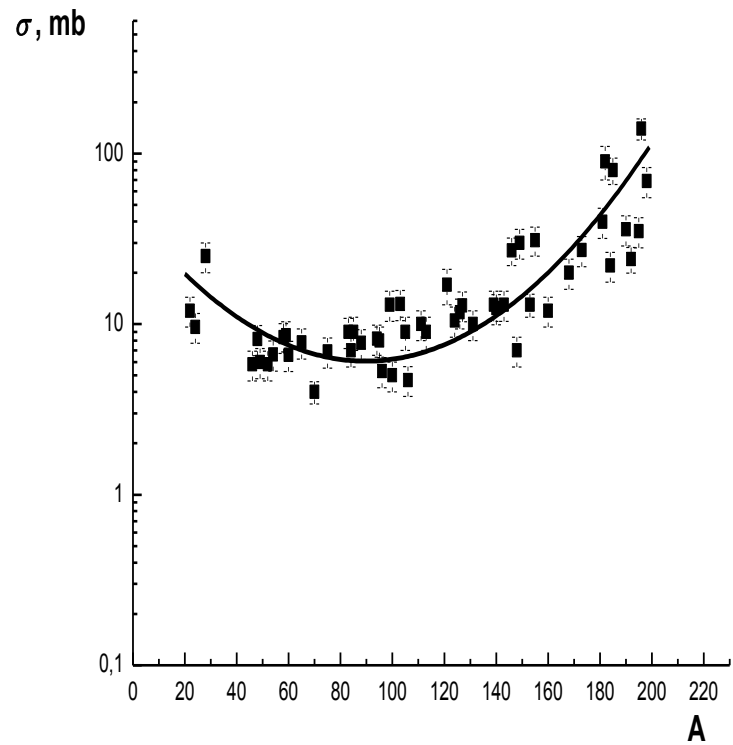}
\caption{\small Mass-yield distribution of 4.4 GeV deuteron-induced
reaction on $^{197}$Au. Solid squares indicate estimated total
isobaric cross sections, solid line is the best fit through the
experimental cross sections of the reactions fragments.}
\end{figure*}

Since the value of impact parameter lies in the range
$1/2(R_p+R_T)\leq b \leq (R_p+R_T)$ \cite{Morrissey}, we can
conclude that the collision on the whole is peripheral. It can be
suggested that in peripheral collisions at large impact parameter
the probability of the interaction of deuteron as whole is very low.
Taking into account the low coupling energy  in deuteron (2.22 MeV)
it can be presented that in this case only one nucleon of deuteron
interacts  with the target. This assumption has been confirmed by
the theoretical work \cite{Cugnon}. The value of $b$ of the
present work is larger than value $b=7.45$ for the system
$^{2}$H+$^{197}$Au at the energy 7.3 GeV \cite{Kozma2} and can
indicate a more peripheral interaction at higher incident energy.

One of the main questions  concerning the reaction  mechanism is the
energy transfer in relativistic nucleus-nucleus collisions. In this
regard, it is interesting to compare our experimental data with the
similar results of the reactions of high energy protons and
deuterons. In Figs. 3 and 4 the ratios of cross section of
residuals from the reactions of 4.4 GeV deuterons on the gold target
of the present work and of 3.65 GeV/nucleon \cite{Kozma3} and 3.0
GeV \cite{Kaufman1} protons with $^{197}$Au, respectively, were
shown. The results can be presented in frames of the two basic
concepts of high-energy nuclear physics, such as factorization and
limiting fragmentation \cite{Feynman}. The first hypothesis assumes
that the yield of particular fragment from the target decay during
the nuclear interactions will not be dependent on the beam except
the geometric factor.

Thus, fragmentation cross sections for the reaction $P+T
\rightarrow F+X$, where $P$ and $T$ are the projectile and target,
and $F$ is the fragment, produced during the reaction, can be
factored according to $\sigma_{T,P}^{F}=\sigma_{T}^{F} \gamma_{p}$,
where $\gamma_{p}$ is dependent only on the projectile. Hence, for
the reactions with protons and deuterons the ratio of cross sections
is

\begin{eqnarray}
\sigma^{F}(d+{}^{197}Au)/ \sigma^{F}(p+{}^{197}Au)=\gamma_{d} / \gamma_{p},
\label{factor1}
\end{eqnarray}
\indent

\noindent which should have constant values $\gamma_{d}/\gamma_{p}$
for any residuals. It means that the ratio should be equal to the
ratio of the total cross sections of reactions.

The ratio of experimental cross sections was determined for target
residues in the mass region from $A=24-196$. The average values of
the relative projectile factors  of our work and data from
\cite{Kozma3} and \cite{Kaufman1} were $\gamma_{d}$/$\gamma_{p}$ =
1.2$\pm$0.2 and 1.55$\pm$0.24, respectively. The ratio of total
reaction cross sections of the present work and data for
\cite{Kozma3} is equal 1.37$\pm$0.34. So it can be supposed that in
the investigated energy range the hypothesis of factorization has
been accomplished.

The hypothesis of limiting fragmentation states that the
distribution of products in the rest frame of the projectile or
target approaches a limiting form as the bombarding energy increases
or, experimentally, that a particular distribution changes
negligibly over a large range of bombarding energies. In order to
test this hypothesis, we compared our results with the deuteron data
on the gold target at 7.3 GeV \cite{Kozma2}. The  ratios of cross
section of residuals are shown in Fig. 5. The average values of the
relative projectile factor in  this case were equal 1.63$\pm$0.24
and total cross section ratio was 0.7$\pm$0.1. From the obtained
results we can conclude that the limiting fragmentation is not valid
for high energy deuterons interactions.

For comparison with model representation of the spallation reactions
induced by protons and deuteron in Au nucleus the intra-nuclear cascade 
and following evaporation for description of the spallation process 
were considered in \cite{Cugnon}. Using the parametrization obtained in
\cite{Cugnon} the mean multiplicity of emitted particles (neutrons
and protons, other charged particles compose a small part) for
deuteron- proton- and neutron-induced reactions on gold target at
the same energy were calculated. The values of these multiplicities
are $<n>=31.121(10\%)$ for $d+^{197}$Au reaction, $<n>=21.03(10\%)$
for $p+^{197}$Au and $<n>=21.33(10\%)$ for $n+^{197}$Au reaction at
the energy 2.2 GeV/nucleon. As we can see, the total neutron
multiplicity in deuteron-induced reactions is less than the sum of
the total neutron multiplicity in proton-induced reactions and of
the same quantity in neutron-induced reactions, for the same target
and the same incident energy per nucleon. The non-additivity of the
cascades initiated by the neutron and the proton contained in the
deuteron comes from two facts. First, for peripheral collisions, one
of the nucleons may not interact at all, corresponding to a
stripping reaction. Second, for more central collisions, the
propagation of one of the nucleons ``clears the space" seen by the
other nucleon.

The ratio of multiplicities of emitted particles in deuteron- and
proton-nuclear reactions is $1.5 \pm 0.2$ should correspond to the
cross section ratio of the residual nuclei formation in average.
This value is in a satisfactory agreement with our experimental
results.

\newpage
\begin{figure*}[h!]
\includegraphics[width=16cm]{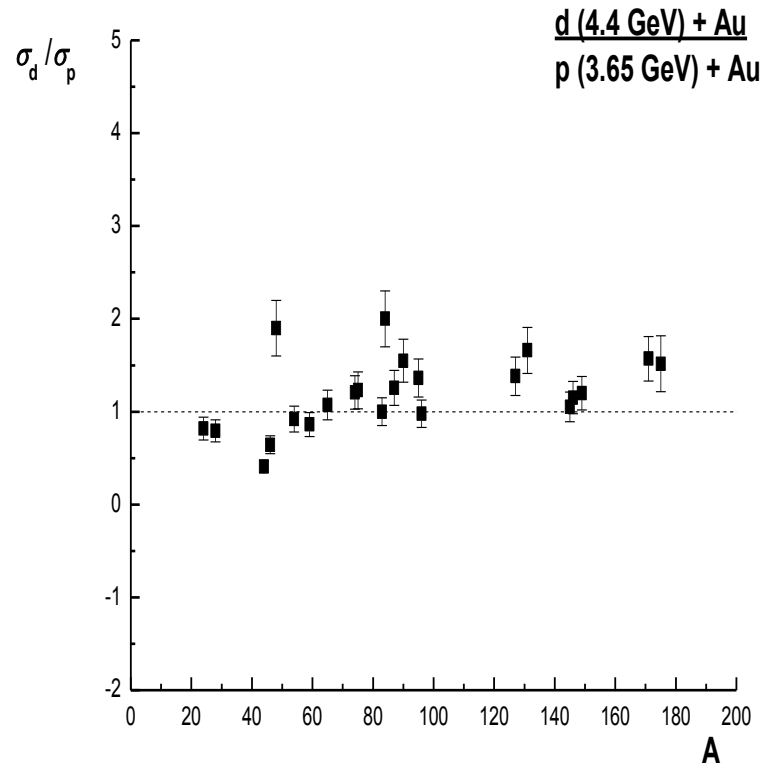}
\caption{\small The ratio of experimental cross sections
$\sigma_{d}/\sigma_{p}$ of  residuals from the reactions of 4.4
GeV deuterons with gold of present work and 3.65 GeV protons
with $^{197}$Au \cite{Kozma3}.}
\end{figure*}

\newpage
\begin{figure*}[h!]
\includegraphics[width=16cm]{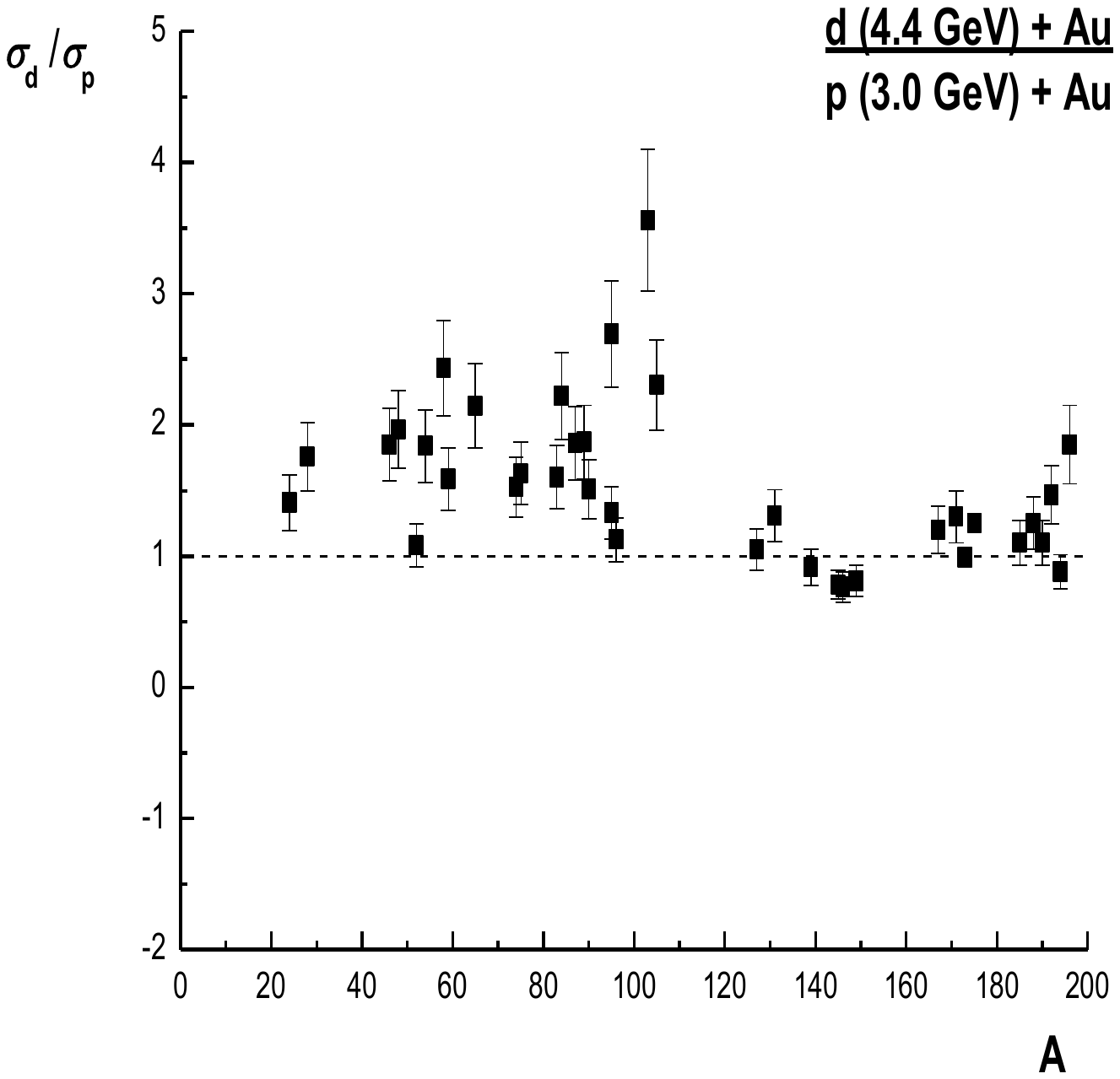}
\caption{\small The ratio of experimental cross sections
$\sigma_{d}/\sigma_{p}$ of residuals from the reactions of 4.4
GeV deuterons with gold of present work and 3.0 GeV protons
with $^{197}$Au \cite{Kaufman1}.}
\end{figure*}

\newpage
\begin{figure*}[h!]
\includegraphics[width=16cm]{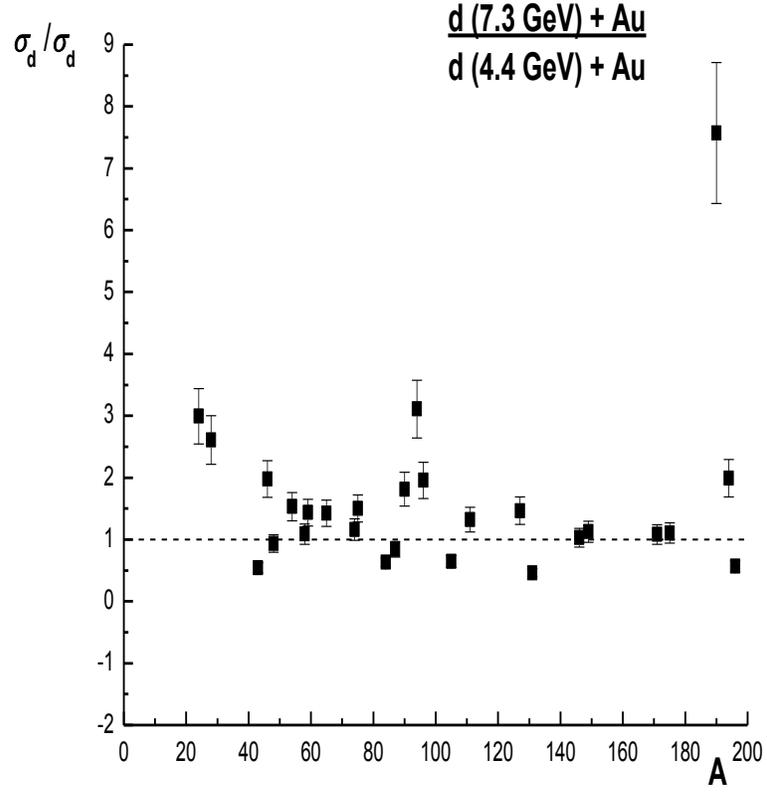}
\caption{\small The ratio of experimental cross sections
$\sigma_{d}/\sigma_{d}$ of  residuals from the reactions of 7.3
GeV deuterons \cite{Kozma2} with gold target and the data
of present work of 4.4 GeV deuterons with $^{197}$Au.}
\end{figure*}

\section*{4 Conclusion}
The cross sections of more than 100 radioactive products formed in
the reaction of deuterons with $^{197}$Au have been measured at
bombarding energy of 4.4 GeV. The charge distribution was analyzed
in the term of 3-parameter equation. It was found that the width of
the charge distribution at a given mass number is the same for all
range of product mass number, but the most probable charge is
smaller for the lighter mass chains. The mass yield distribution of
target residues has been determined by integration of the cross
sections for fragments with $A > 40$ and has been compared with
the theoretical predictions. It was suggested that a few sources
having different excitation energies can take part in the formation
of residual nuclei, fission, production of the IMFs and spallation.

The calculation of the impact parameter indicates
that target residues from the reactions of 4.4 GeV deuterons
with gold are created in peripheral collisions.

The ratio of the experimental cross sections of proton and
deuteron induced reactions is in a good agreement with the
ratio of theoretical calculations of the multiplicity of the
emitted particles in the cascade and evaporation stages.

The mechanism of interaction of deuteron with nucleus is not
provided by the additional interaction of two nucleons in deuteron
in the interaction with nuclei. The deuteron as weak bound nucleus
can decay and demonstrate one nucleon collision in most cases of
interactions. The interaction was of peripheral character in most
cases as it was estimated from the calculation of the impact
parameter of deuteron-nucleus interaction.

The similarity between the cross section of target residues of this
work and those from the reactions with proton- and deuteron-induced
reactions on the gold target at the different kinetic energies may
be viewed as an evidence for factorization in these reactions.
The condition of the limiting fragmentation is not valid for
the energy deuterons up to 7.3 GeV.

The ratio of the cross sections of proton- and deuteron-induced reactions
can confirm the fulfilling of the factorization conditions in the
investigated energy range.

\section*{Acknowledgment}
G. Karapetyan is grateful to Funda\c c\~ao de Amparo \`a Pesquisa do
Estado de S\~ao  Paulo (FAPESP) 2011/00314-0 and
to International Centre for Theoretical Physics (ICTP) under
the Associate Grant Scheme.

The authors are grateful to group leader of LNP JINR Dr. S. Avdeev
for granting possibility of carrying out experiment, and also to
lead researcher of LHEP JINR Kh. Abraamyan for the help during the
experiment.

\medbreak
\end{document}